\documentclass[a4paper,pra,footinbib,reprint,twocolumn,superscriptaddress]{revtex4}

\usepackage{ulem}
\usepackage{amssymb}
\usepackage{amsmath}
\usepackage{epsfig}
\usepackage{color}
\usepackage{graphics, graphicx}
\usepackage{bbold}
\usepackage{psfrag}
\usepackage{mathcomp}
\usepackage{amsmath}
\usepackage{amssymb}%
\usepackage{mathrsfs}%
\usepackage{subfigure}
\usepackage{verbatim}
\usepackage{xcolor}
\usepackage[colorlinks,citecolor=blue,urlcolor=blue]{hyperref}

\begin{document}

\date{\today}
\title{Superfluid phases and excitations in a cold gas of  $d$-wave interacting bosonic atoms and molecules}
\author{Zehan Li}
\affiliation{Department of Physics and Astronomy, University of Pittsburgh, Pittsburgh, PA 15260, USA}
\author{Jian-Song Pan\footnote{Current address: Department of Physics, National University of Singapore, Singapore 117543, Singapore.}}
\email{panjsong@gmail.com}
\affiliation{Wilczek Quantum Center, School of Physics and Astronomy and T. D. Lee Institute,
		Shanghai Jiao Tong University, Shanghai 200240, China}
\author{W. Vincent Liu}
\email{wvliu@pitt.edu}
\affiliation{Department of Physics and Astronomy, University of Pittsburgh, Pittsburgh, PA 15260, USA}
\affiliation{Wilczek Quantum Center, School of Physics and Astronomy and T. D. Lee Institute, Shanghai Jiao Tong University, Shanghai 200240, China}
	
\begin{abstract}
Motivated by recent advance in orbitally tuned Feshbach resonance experiments, we analyze the ground-state phase diagram and related low-energy excitation spectra of a $d$-wave interacting Bose gas. A two-channel model with $d$-wave symmetric interactions and background $s$-wave interactions is adopted to characterize the gas. The ground state is found to show three interesting phases: atomic, molecular, and atomic-molecular superfluidity. Remarkably differently from what was previously known in the $p$-wave case, the atomic superfluid is found to be momentum-independent in the present $d$-wave case. Bogoliubov spectra above each superfluid phase are obtained both analytically and numerically.
\end{abstract}
\pacs{67.85.Lm, 03.75.Ss, 05.30.Fk}

\maketitle

\section{Introduction}
Orbitally high-partial-wave interacting quantum gases~\cite{chin2010feshbach} steadily attract research interest due to the potential to show exotic superfluididty. For example, $d$-wave interacting Fermi gases may be employed to compare the $d$-wave superfluid. Recently, $d$-wave scattering resonance was observed in more and more ultracold atomic gases~\cite{werner2005observation,beaufils2009feshbach,cui2017observation,yao2019degenerate, zhu2019high}. Particularly the observation of degenerate $d$-wave-interacting Bose gases with $d$-wave shape resonance~\cite{yao2019degenerate} makes the hidden $d$-wave many-body correlation experimentally more accessible.

Unlike s-wave interaction, the closed channels of high-partial-wave Feshbach resonance carry finite momentum. For example, the closed channels of the $p$-wave Feshbach resonance carry a total angular momentum of $1\hbar$ and the interaction term is proportional to momentum $k$. It is predicted that finite-momentum superfluid emerges in a $p$-wave interacting Bose gas~\cite{radzihovsky2009p,pwave2011Sungsoo,spontaneous2019li}. The closed channels of $d$-wave Feshbach resonance carry a total angular momentum of $2\hbar$, and hence the many-body form is proportional to the square of momentum $k^2$.  Although $d$-wave electronic Fermi superconductor has been studied extensively in condensed matter physics, to the best of our knowledge, what possible many-body states the $d$-wave interacting atomic Bose gases should exhibit is a widely open question.

Inspired by recent experimental progress~\cite{cui2017observation,yao2019degenerate, zhu2019high}, we analyze the zero-temperature mean-field ground state and Bogoliubov spectrum of a $d$-wave interacting Bose gas in this paper. A two-channel model is adopted for a mixture of two components interacting via $d$-wave interaction. Similar to the $p$-wave interacting Bose gas~\cite{radzihovsky2009p,pwave2011Sungsoo,spontaneous2019li}, the mean-field ground state typically shows three quantum phases: atomic superfluid (ASF), molecular superfluid (MSF) and atomic-molecular superfluid (AMSF). But unlike the $p$-wave case, the atomic superfluid does not carry finite momentum. The phase boundaries are analytically obtained. Further, the Bogoliubov excitation spectrum is analyzed both numerically and analytically above the superfluid groundstate with d-orbital aspects.

\section{Model}
Inspired by the experiments\cite{cui2017observation, yao2017strongly}, we will focus on a gas mixture of two distinguishable bosonic atoms (e.g., $^{85}$Rb and $^{87}$Rb). The two atomic fields are created by $\hat{\psi}_\sigma^\dagger=(\hat{\psi}_1^\dagger, \hat{\psi}_2^\dagger)$ and interact through a $d$-wave FR associated with a tunable molecular bound state~\cite{pwave2011Sungsoo, dwave2018juanyao}. According to the symmetry of this system, the angular momentum is a good quantum number and the related $d$-wave molecule (e.g., $^{85}$Rb-$^{87}$Rb) field is created by $\hat{\phi}_m^\dagger=(\hat{\phi}_{-2}^\dagger, \hat{\phi}_{-1}^\dagger, \hat{\phi}_{0}^\dagger, \hat{\phi}_{1}^\dagger, \hat{\phi}_{2}^\dagger)$, which corresponds to the five closed-channel molecule states (e.g., $l_z$=0, $\pm1$, $\pm2$). The Hamiltonian density for this system is written as (we take $\hbar$=1 throughout) \cite{dwave2016Peifeng},
\begin{equation}\label{eq_mainH}
\begin{split}
\mathcal{H}=&
\sum_{\sigma=1,2}\hat{\psi}_\sigma^\dagger(-\frac{\nabla^2}{2m}-\mu_\sigma)\hat{\psi}_\sigma+\sum_{m=-2}^{2}[\hat{\phi}_m^\dagger(-\frac{\nabla^2}{4m}+z(-\frac{\nabla^2}{4m})^2\\
&-\mu_M)\hat{\phi}_m.
 -\overline{g}(\hat{\phi}_m^\dagger y_m+h.c.)]+\mathcal{H}_{bg},
\end{split}
\end{equation}
where $y_m$~\cite{dwave2016Peifeng} and $\mathcal{H}_{bg}$ are respectively given by
\begin{equation}
\begin{split}
y_m=&\frac{1}{4}\sum\limits_{a,b=x,y,z}C^m_{ab}[(\partial_a \hat{\psi}_1) (\partial_b \hat{\psi}_2) - (\partial_a \partial_b \hat{\psi}_1) \hat{\psi}_2\\
&+(\partial_b \hat{\psi}_1) (\partial_a \hat{\psi}_2) - \hat{\psi}_1 (\partial_a \partial_b \hat{\psi}_2)],
\end{split}
\end{equation}
\begin{equation}
\begin{split}
\mathcal{H}_{bg}=&\frac{1}{2}\sum_{\sigma,\sigma^{'}=1,2}\lambda_{\sigma\sigma^{'}}|\hat{\psi}_{\sigma}|^{2}|\hat{\psi}_{\sigma^{'}}|^{2}+\sum_{m,n=-2}^2\frac{g_0}{2}(\hat{\phi}_m^\dagger \hat{\phi}_m)(\hat{\phi}_n^\dagger \hat{\phi}_n)\\
&+\sum_{m=-2}^{2}g_\mathrm{AM}({|\hat{\psi}_1|}^2+{|\hat{\psi}_2|}^2)\hat{\phi}_m^\dagger \hat{\phi}_m.
\end{split}
\end{equation}
Here $\mu_1$ and $\mu_2$ are the chemical potentials of the atoms and $\mu_M$ is that for molecule.  The detuning between atomic and molecular channels is given by $\nu=\mu_1+\mu_2-\mu_M$. $\overline{g}$ characterizes the $d$-wave interaction strength. $C^m_{ab}$ is the Clebsch-Gordan coefficient~\cite{dwave2018juanyao}, satisfying $\sum_{a,b}C^m_{ab}k_a k_b/k^2=\sqrt{4\pi}Y_2^m(\hat{k})$, where $Y_2^m(\hat{k})$ is the spherical harmonics. For simplicity, we have taken the two atomic masses $m$ to be identical, which is a good approximation for the $^{85}$Rb and $^{87}$Rb mixture. In the background interaction $\mathcal{H}_{bg}$, the $\lambda_{\sigma\sigma^{'}}$ term characterizes the atom-atom interactions given by different species respectively, the $g_\mathrm{AM}$ term describes the atom-molecule interaction, and the $g_0$ term describes the molecule-molecule interaction.

\section{Mean-Field Theory}
We will obtain the Landau free energy by applying mean-field theory to our model and minimize it to establish the phase diagram and analyze the phase transition. This method is equivalent to solving Gross-Pitaevskii equation. Replacing the atomic and molecular field operators with their relative classical order parameters $\Psi_\sigma, \Phi_m$, we obtain the Landau free energy function $F[\Psi_\sigma, \Phi_m]=\langle H \rangle$.

\begin{figure}[tbp]
	\includegraphics[width=7.5cm]{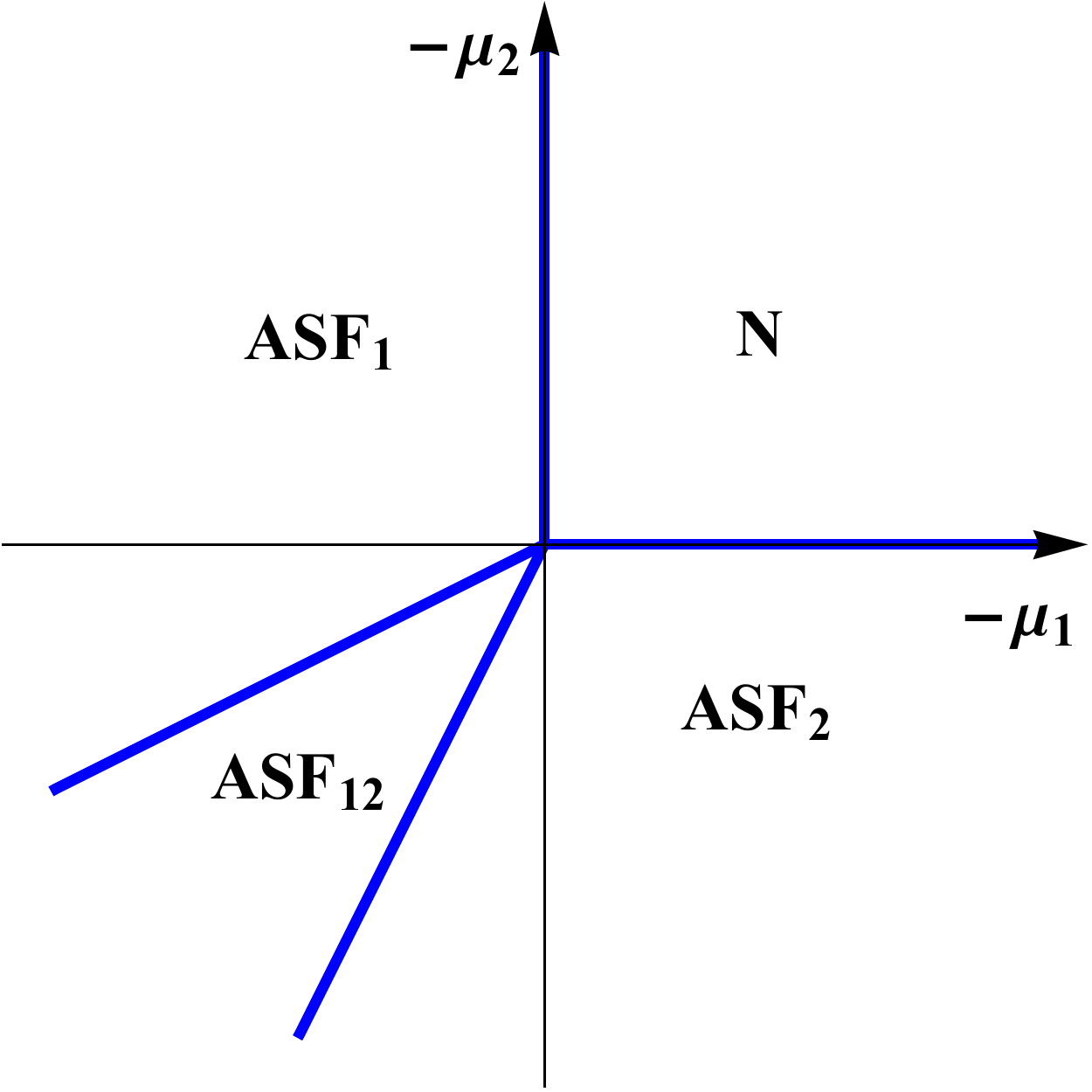}
	\caption{Mean-field phase diagram of a $d$-wave resonant two-component Bose gas for large positive detuning and $4\lambda_{11} \lambda_{22}-(\lambda_{12}+\lambda_{21})^2>0$. The atomic channels have lower energy. $ASF_1$ and $ASF_2$ refer to single atom species superfluid state, and $ASF_{12}$ refers to double atom species superfluid state.}
	\label{fig:model}
\end{figure}

We decompose our mean-field parameters to characterize the states of the system. For the atomic condensates $\Psi_1$ and $\Psi_2$, let us use Fourier transform and make these fields complex periodic functions characterized by momenta $\textbf{Q}_n$,
\begin{eqnarray}
\Psi_\sigma = \sum_{\textbf{Q}_n}\Psi_{\sigma,\textbf{Q}_n}e^{i\textbf{Q}_n\cdot\textbf{r}}.
\end{eqnarray}
\noindent
It is generally expected that the assumption of having a single component, $\textbf{Q}_n=\textbf{Q}$ is sufficient to capture the qualitative picture of the ground state~\cite{finiteQ2006kuklov}. Based on the energetics of the model (see supplementary), the simplest single $\textbf{Q}_1=\textbf{Q}$ form is given by
\begin{eqnarray}\label{eq_ansatz}
\Psi_1 &= \Psi_{1,\textbf{Q}}e^{i\textbf{Q}\cdot\textbf{r}},\\ \nonumber
\Psi_2 &= \Psi_{2,-\textbf{Q}}e^{-i\textbf{Q}\cdot\textbf{r}}.
\end{eqnarray}

\subsection{Atomic Superfluid Phase}
For large positive detuning $\nu>0$, the atomic channels have lower energy and the ground state is a molecule vacuum. Thus the molecular order parameters vanish, leading to an effective atomic free energy~\cite{pwave2011Sungsoo},
\begin{eqnarray}
F_A[\Psi_\sigma]&=&\left.
\int d^3r\sum_{\sigma=1,2}\Psi_\sigma^\ast(-\frac{\nabla^2}{2m}-\mu_\sigma)\Psi_\sigma\right.\\
&& \left.\nonumber
+
\sum_{\sigma,\sigma'=1,2}\frac{\lambda_{\sigma,\sigma'}}{2}{|\Psi_\sigma|}^2{|\Psi_{\sigma'}|}^2\right..
\end{eqnarray}
This free energy is minimized by spatially uniform atomic order parameters~\cite{swave2003calabrese} and leads to the free energy density $f_A=F_A/V$ with the form
\begin{eqnarray}
f_A&=&\left.
-\sum_{\sigma=1,2}\mu_\sigma{|\Psi_\sigma|}^2
+\sum_{\sigma,\sigma'=1,2}\frac{\lambda_{\sigma,\sigma'}}{2}{|\Psi_\sigma|}^2{|\Psi_{\sigma'}|}^2\right..
\end{eqnarray}
\noindent
For  $4\lambda_{11} \lambda_{22}-(\lambda_{12}+\lambda_{21})^2>0$, the minimization of $f_A$ leads to different superfluid phases as $\mu_1$ and $\mu_2$ change, which are listed in Table~\ref{tab:mean_field_phases} (see Fig. 1). Otherwise, for $4\lambda_{11} \lambda_{22}-(\lambda_{12}+\lambda_{21})^2<0$, the ${\text{ASF}}_{\text{12}}$ phase tends to be unstable, there will be a direct first-order phase transition from ${\text{ASF}}_{\text{1}}$ to ${\text{ASF}}_{\text{2}}$, and its phase boundary is determined to be $\mu_2=\sqrt{\frac{\lambda_{22}}{\lambda_{11}}}\mu_1$ (see Fig. 2).

\begin{table*}
	\begin{tabular}{|c|c|c|c|c|}
		\hline
		phase & chemical potentials & $\Psi_1$ & $\Psi_2$\tabularnewline
		\hline
		$N$ & $\mu_1<0,\mu_2<0$ & 0 & 0\tabularnewline
		\hline
		$ASF_1$ & $\mu_1>0,\mu_2<\frac{\lambda_{12}+\lambda_{21}}{2\lambda_{11}}\mu_1$ & $\sqrt{\frac{\mu_1}{\lambda_{11}}}$ & 0\tabularnewline
		\hline
		$ASF_2$ & $\mu_1<\frac{\lambda_{12}+\lambda_{21}}{2\lambda_{22}}\mu_2,\mu_2>0$ & 0 & $\sqrt{\frac{\mu_2}{\lambda_{22}}}$\tabularnewline
		\hline
		$ASF_{12}$ & $\mu_1>\frac{\lambda_{12}+\lambda_{21}}{2\lambda_{22}}\mu_2, \mu_2>\frac{\lambda_{12}+\lambda_{21}}{2\lambda_{11}}\mu_1$ & $\sqrt{\frac{4\lambda_{22} \mu_1-2(\lambda_{12}+\lambda_{21}) \mu_2}{4\lambda_{11} \lambda_{22}-(\lambda_{12}+\lambda_{21})^2}}$ & $\sqrt{\frac{4\lambda_{11} \mu_2-2(\lambda_{12}+\lambda_{21}) \mu_1}{4\lambda_{11} \lambda_{22}-(\lambda_{12}+\lambda_{21})^2}}$\tabularnewline
		\hline
	\end{tabular}
	\caption{Sub-phases of the ASF phase. i) When $\mu_1$ and $\mu_2$ are negative, both atomic species are in the normal (N) phase. ii) When $\mu_1>0,\mu_2<\frac{\lambda_{12}+\lambda_{21}}{2\lambda_{11}}\mu_1$, the atom 1 forms condensate. iii) When $\mu_1<\frac{\lambda_{12}+\lambda_{21}}{2\lambda_{22}}\mu_2,\mu_2>0$, the atom 2 forms condensate. iv) When $\mu_1>\frac{\lambda_{12}+\lambda_{21}}{2\lambda_{22}}\mu_2, \mu_2>\frac{\lambda_{12}+\lambda_{21}}{2\lambda_{11}}\mu_1$, both atom species form condensates.}\label{tab:mean_field_phases}
\end{table*}

\subsection{Molecular Superfluid Phase}
In the MSF phase, we have large negative detuning $\nu<0$, that is, $-\nu \gg |\mu_{1,2}|$. The molecular channels have lower energy and the ground state is an atom vacuum. The free energy density $f_M$ is given as
\small
\begin{eqnarray}
f_M&=&\left.
\sum_{m=-2}^{2}-\mu_M|\Phi_m|^2
+
\sum_{m,n=-2}^{2}\frac{g_0}{2} (\Phi_m^\ast \Phi_m)(\Phi_n^\ast \Phi_n). \right.
\end{eqnarray}
\noindent
The molecular condensate density is obtained by minimizing the free energy,
\noindent
\begin{eqnarray}
\newcommand{\colvec}[2][.8]{%
	\scalebox{#1}{%
		\renewcommand{\arraystretch}{.8}%
		$\begin{bmatrix}#2\end{bmatrix}$%
	}
}
\begin{split}\label{eq_GSmol}
\Phi&=\sqrt{\frac{\mu_M}{g_0}}D(0,0,1,0,0)^T,
\end{split}
\end{eqnarray}
\noindent
where $D$ is an $SU(5)$ matrix satisfying $D*D^\dagger=1$. The ground state implies a broken symmetry group $SU(5)$.

\subsection{Atomic-Molecular Superfluid}
For the intermediate detuning, both the atomic and molecular modes are gapless. To understand the phase boundaries and the behavior of order parameters, it is convenient to approach the AMSF phase from MSF phase~\cite{pwave2011Sungsoo}. For simplicity, we specialize in a balanced mixture by $\mu_1=\mu_2=\mu$. Applying mean-field assumption, we obtain the free energy density $f_{AM}=F[\Psi_\sigma,\Phi_m]/V=f_Q+f_M$, where $f_Q$ describes the $Q$-dependent fragment in the free energy density $f_{AM}$,
\begin{eqnarray}\label{eq_fQ}
f_Q&=&\left.
\sum_{\sigma=1,2}\varepsilon_Q|\Psi_{\sigma,\textbf{Q}_\sigma}|^2-(\Delta_{\textbf{Q}}^\ast \Psi_{1,\textbf{Q}} \Psi_{2,-\textbf{Q}}
+
c.c.)\right.\\ \nonumber
&& \left.
+
\sum_{\sigma,\sigma'=1,2}\frac{\lambda_{\sigma,\sigma'}}{2}{|\Psi_{\sigma,\textbf{Q}_\sigma}|}^2{|\Psi_{\sigma',\textbf{Q}_{\sigma'}}|}^2
\right.,
\end{eqnarray}
\begin{eqnarray}\label{eq_fM}
f_M&=&\left.
-\sum_{m=-2}^{2}\mu_M\Phi_m^\ast\Phi_m+\sum_{m,n=-2}^2\frac{g_0}{2}(\Phi_m^\ast \Phi_m)(\Phi_n^\ast \Phi_n).\right.
\end{eqnarray}
\noindent
where the atomic order parameter ansatz Eqs.~(\ref{eq_ansatz}) is used to simplify $f_Q$. $\varepsilon_Q=(\frac{Q^2}{2m}-\mu+\sum_{m=-2}^{2}g_\mathrm{AM}|\Phi_m|^2), \Delta_{\textbf{Q}}=\sum_{m=-2}^{2}\overline{g} \sqrt{4\pi} Q^2 Y_{2}^m(\hat{\textbf{Q}}) \Phi_m, \textbf{Q}_1=\textbf{Q}$ and $\textbf{Q}_2=-\textbf{Q}$. When we approach the ASMF phase from the MSF phase, the atomic condensate fractions are considered to be small and perturbative. Thus, in Eq.~(\ref{eq_fQ}), the quadratic order terms are enough to characterize the free energy density $f_Q$. We focus on these terms for the time being,
\begin{equation}
\begin{split}
f_Q^0&=\label{eq}
\sum_{\sigma=1,2}\varepsilon_Q|\Psi_{\sigma,\textbf{Q}_\sigma}|^2-(\Delta_{\textbf{Q}}^\ast \Psi_{1,\textbf{Q}} \Psi_{2,-\textbf{Q}}
+
c.c.)\\
&=(\Psi_{1,\textbf{Q}}^\ast \quad \Psi_{2, -\textbf{Q}}) \begin{pmatrix} \varepsilon_Q & -\Delta_{\textbf{Q}} \\ -\Delta_{\textbf{Q}}^\ast & \varepsilon_Q \end{pmatrix} \begin{pmatrix} \Psi_{1,\textbf{Q}} \\ \Psi_{2, -\textbf{Q}}^\ast \end{pmatrix}\\
&=\epsilon_Q^+ |\Psi_+|^2+\epsilon_Q^- |\Psi_-|^2,
\end{split}
\end{equation}
\\*
where the free energy density is written in a diagonalized formula. The eigenvalues $\epsilon_Q^+, \epsilon_Q^-$ and eigenvectors $\Psi_-, \Psi_+^\ast$ are listed below,
\begin{eqnarray}
\begin{pmatrix} \epsilon_Q^+ \\ \epsilon_Q^- \end{pmatrix}=\begin{pmatrix} \varepsilon_Q+|\Delta_{\textbf{Q}}| \\ \varepsilon_Q-|\Delta_{\textbf{Q}}| \end{pmatrix}.
\end{eqnarray}
\begin{eqnarray}\label{eq_Psipm}
\begin{pmatrix} \Psi_- \\ \Psi_+^\ast \end{pmatrix}=\frac{1}{\sqrt{2}}\begin{pmatrix} e^{-i\theta_0}\Psi_{1,\textbf{Q}}+\Psi_{2,-\textbf{Q}}^\ast \\ -e^{-i\theta_0}\Psi_{1,\textbf{Q}}+\Psi_{2,-\textbf{Q}}^\ast \end{pmatrix},
\end{eqnarray}
\noindent
where $\theta_0$ is the angle of $\Delta_{\textbf{Q}}$, $\Delta_{\textbf{Q}}=|\Delta_{\textbf{Q}}|e^{i\theta_0}$.
When the atom condensate is emergent in the AMSF phase, they prefer to stay at a lower energy level, which means $\Psi_+=0$. We deduce that
\begin{eqnarray}\label{eq_GSatom}
\Psi_{2,-\textbf{Q}}^\ast=e^{-i\theta_0}\Psi_{1,\textbf{Q}}.
\end{eqnarray}
\\*
Furthermore, a zero momentum solution is obtained to minimize the $Q$-dependent fragment of the free energy,
\begin{eqnarray}\label{eq_Q0}
Q=0,
\end{eqnarray}
\begin{eqnarray}\label{eq_GSPsim}
|\Psi_-|=\sqrt{\frac{-\epsilon_Q^-}{\lambda}},
\end{eqnarray}
where $\epsilon_Q^-=-\mu+\sum_{m=-2}^{2}g_\mathrm{AM}\Phi_m^\ast\Phi_m$, $\lambda=\frac{1}{4}(\lambda_{11}+\lambda_{22}+\lambda_{12}+\lambda_{21})$. The above result Eq.~(\ref{eq_Q0}) implies that unlike the $p$-wave case~\cite{radzihovsky2009p}, the atomic superfluid is shown to be momentum-independent.

\begin{figure}[tbp]
	\includegraphics[width=7.5cm]{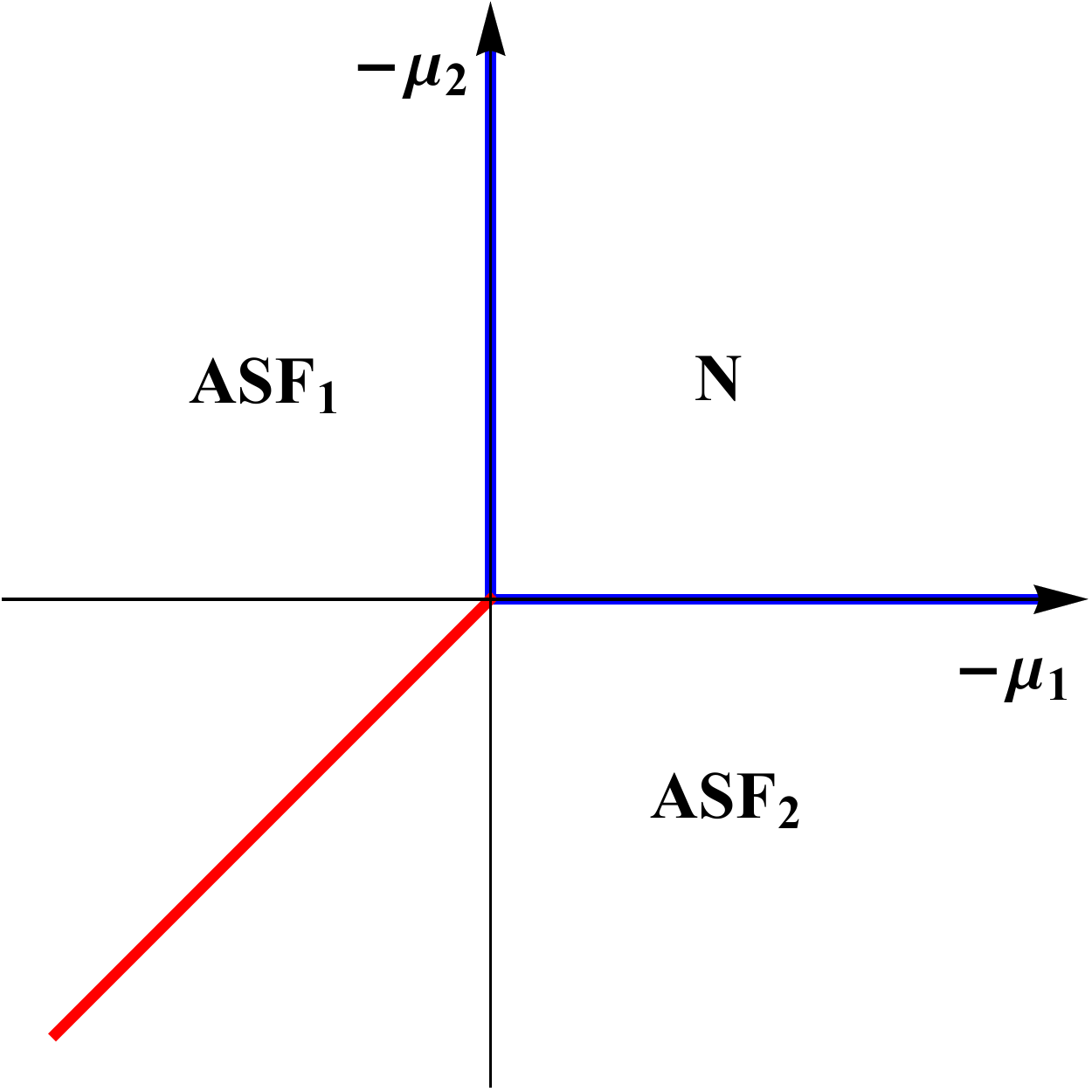}
	\caption{Mean-field phase diagram of a $d$-wave resonant two-component Bose gas for large positive detuning and $4\lambda_{11} \lambda_{22}-(\lambda_{12}+\lambda_{21})^2<0$. A valid phase of significant condensate fraction in both atom fields is not found in mean-field calculation. The phases $ASF_1$ and $ASF_2$ are separated by a first-order transition boundary.}
	\label{fig:model}
\end{figure}

\begin{figure}[tbp]
	\includegraphics[width=7.5cm]{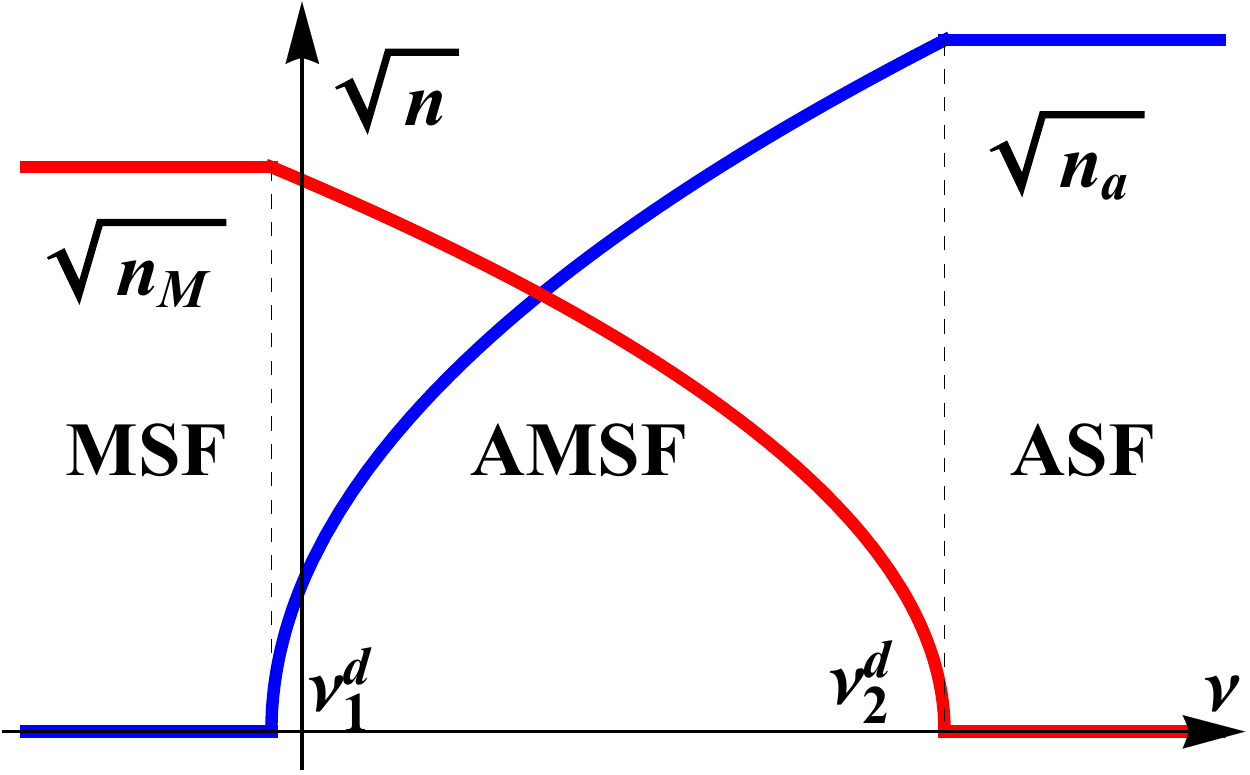}
	\caption{Schematic atomic and molecular condensate density versus the FR detuning $\nu$. Red curves are for molecule condensate density, blue curves are for atom condensate density, i) MSF for $\nu<\nu_1^d$ ii) AMSF for $\nu_1^d<\nu<\nu_2^d$ iii) ASF for $\nu>\nu_2^d$.}
	\label{fig:model}
\end{figure}

Now let us come back to $f_{AM}$ by adding molecular related energy density into Eq.~(\ref{eq_fQ}), and simplify it with Eq.~(\ref{eq_GSatom}),
\begin{eqnarray}
f_{AM}&=&\left.\nonumber-\frac{1}{2 \lambda}[\mu-g_\mathrm{AM}\sum_{m=-2}^{2}|\Phi_m|^2]^2
-\mu_M\sum_{m=-2}^{2}|\Phi_m|^2\right.\\
&& \left.
+
\frac{g_0}{2}\sum_{m,n=-2}^{2} (\Phi_m^\ast \Phi_m)(\Phi_n^\ast \Phi_n)\right..
\end{eqnarray}
\noindent
The condensate mean-field ground states are obtained to minimize the free energy $f_{AM}$,
\begin{eqnarray}
\begin{split}
\Phi&=\sqrt{\frac{g_\mathrm{AM}\mu-\lambda\mu_M}{g_\mathrm{AM}^2-g_0\lambda}}D (0,0,1,0,0)^T,
\end{split}
\end{eqnarray}
\begin{eqnarray}
|\Psi_{1,2}|=\sqrt{\frac{g_0\mu-g_\mathrm{AM}\mu_M}{2\lambda g_0-2g_\mathrm{AM}^2}},
\end{eqnarray}
\noindent
where $D$ is an $SU(5)$ rotation matrix. Similar to the analysis in the MSF context, the broken symmetry group is $SU(5)$. The condensate densities are
\begin{eqnarray}
n_M=\frac{(2\lambda-g_\mathrm{AM})\mu-\lambda \nu}{\lambda g_0-g_\mathrm{AM}^2},
\end{eqnarray}
\begin{eqnarray}
n_A=\frac{(g_0-2g_\mathrm{AM})\mu+g_\mathrm{AM}\nu}{\lambda g_0-g_\mathrm{AM}^2}.
\end{eqnarray}
\noindent
By setting $n_A=0$ and $n_M=0$ respectively, we obtain the two phase boundaries to separate the three phases molecular superfluid (MSF), atomic-molecular superfluid (AMSF) and atomic superfluid (ASF):
\begin{eqnarray}
\nu_1^d=(2-\frac{g_0}{g_\mathrm{AM}})\mu,\label{eq_nu1d},\\
\nu_2^d=(2-\frac{g_\mathrm{AM}}{\lambda})\mu.\label{eq_nu2d}
\end{eqnarray}

\section{Low Energy Excitations}
In this section, we will focus on the low energy excitations for $d$-wave FR to cross  examine the consistency of mean-field results. To begin with, we expand the field operators in the ASF, MSF and AMSF phases around their mean-field condensate values~\cite{spontaneous2019li, pwave2011Sungsoo},
\begin{equation}
\begin{split}
&\hat{\psi}_{\sigma}=\Psi_\sigma+\delta \hat{\psi}_{\sigma},\\
&\hat{\phi}_m=\Phi_m+\delta \hat{\phi}_m,
\end{split}
\end{equation}
in the momentum space,
\begin{eqnarray}
\delta \hat{\psi}_{\sigma}=\frac{1}{\sqrt{V}}\sum_{\textbf{k}}\hat{a}_{\sigma, \textbf{k}}e^{i\textbf{k} \cdot \textbf{r}},\\ \nonumber
\delta \hat{\phi}_{m}=\frac{1}{\sqrt{V}}\sum_{\textbf{k}}\hat{b}_{m, \textbf{k}}e^{i\textbf{k} \cdot \textbf{r}}.
\end{eqnarray}
With the above representations, the Hamiltonian~(\ref{eq_mainH}) is expanded up to the second order in terms of the operators $\hat{a}_{\sigma, \textbf{k}}$ and $\hat{b}_{m, \textbf{k}}$,
\begin{widetext}
	\begin{eqnarray}\label{BogoH}
	H_f=&&\sum_{\textbf{k}}\{\sum_{\sigma=1,2}(\frac{1}{2}\varepsilon_{\sigma,\textbf{k}+\textbf{Q}_\sigma}\hat{a}_{\sigma,\textbf{k}+\textbf{Q}_\sigma}^\dagger \hat{a}_{\sigma,\textbf{k}+\textbf{Q}_\sigma}+\tilde{\lambda}_{\sigma} \hat{a}_{\sigma,-\textbf{k}+\textbf{Q}_\sigma}\hat{a}_{\sigma,\textbf{k}+\textbf{Q}_\sigma})
	+
	t_1 \hat{a}_{1,\textbf{k}+\textbf{Q}}^\dagger \hat{a}_{2,\textbf{k}-\textbf{Q}}
	+
	t_{2,\textbf{k}+\textbf{Q}} \hat{a}_{1,\textbf{k}+\textbf{Q}} \hat{a}_{2,-\textbf{k}-\textbf{Q}}\\\nonumber
	+
	&&\sum_{m} (\frac{1}{2}\omega_{m,k} \hat{b}_{m,k}^\dagger \hat{b}_{m,k}+\delta_m \hat{b}_{m,-k} \hat{b}_{m,k})
	+
	\frac{1}{2}\sum_{m\neq n} (g_{m,n} \hat{b}_{n,k}^\dagger \hat{b}_{m,k}+\gamma_{m,n} \hat{b}_{m,-k} \hat{b}_{n,k})\\\nonumber
	+
	&&\sum_{\sigma,m} \beta_{1,m,\sigma} \hat{a}_{\sigma, \textbf{k}+\textbf{Q}_\sigma}^\dagger \hat{b}_{m,k} + \sum_{\sigma,m} \beta_{2,m,\sigma} \hat{a}_{\sigma, \textbf{k}+\textbf{Q}_\sigma}^\dagger \hat{b}_{m,-k}^\dagger + \sum_{\sigma,m} \beta_{3,m,\sigma} \hat{a}_{\sigma, -\textbf{k}+\textbf{Q}_\sigma} \hat{b}_{m,k} + \sum_{\sigma,m} \beta_{4,m,\sigma} \hat{a}_{\sigma, -\textbf{k}+\textbf{Q}_\sigma} \hat{b}_{m,-k}^\dagger\\\nonumber
	-
	&&\sum_{\sigma,m} \alpha_{m,\overline{\sigma},\textbf{k}} \hat{b}_{m,k}^\dagger \hat{a}_{\sigma, \textbf{k}+\textbf{Q}_\sigma} + h.c.
	\}.
	\end{eqnarray}
\end{widetext}

The parameters are defined below,
\begin{eqnarray}
\varepsilon_{\sigma,\textbf{k}}&=&\left.\epsilon_{\textbf{k}}-\mu_\sigma+2\lambda_{\sigma,\sigma}|\Psi_\sigma|^2+\frac{1}{2}(\lambda_{12}+\lambda_{21})|\Psi_{\overline{\sigma}}|^2\right.\\\nonumber
&& \left.
+
g_\mathrm{AM}\sum_{m}\Phi_m^\ast\Phi_m\right.,
\end{eqnarray}
\begin{eqnarray}
\omega_{m,k}&=&\left.\frac{1}{2}\epsilon_{\textbf{k}}+z(\frac{1}{2}\epsilon_{\textbf{k}})^2-\mu_M+g_{0}\sum_{n}\Phi_n^\ast\Phi_n\right.\\
&& \left.\nonumber
+
g_{0}\Phi_m^\ast\Phi_m+g_\mathrm{AM}(|\Psi_1|^2+|\Psi_2|^2)\right.,
\end{eqnarray}
\begin{eqnarray}
\tilde{\lambda}_\sigma=\frac{1}{2}\lambda_{\sigma,\sigma}\Psi_\sigma^{\ast2},
\end{eqnarray}
\begin{eqnarray}
t_1=\frac{1}{2}(\lambda_{12}+\lambda_{21})\Psi_1 \Psi_2^\ast,
\end{eqnarray}
\begin{eqnarray}
t_{2,\textbf{k}}=\frac{1}{2}(\lambda_{12}+\lambda_{21})\Psi_1^\ast \Psi_2^\ast-\overline{g}\sqrt{4\pi}k^2 \sum_{m}\Phi_m^\ast Y_2^m(\hat{\textbf{k}}),
\end{eqnarray}
\begin{eqnarray}
\delta_m=\frac{1}{2}g_0\Phi_m^\ast \Phi_m^\ast,
\end{eqnarray}
\begin{eqnarray}
g_{m,n}=g_0\Phi_m^\ast \Phi_n,
\end{eqnarray}
\begin{eqnarray}
\gamma_{m,n}=g_0\Phi_m^\ast \Phi_n^\ast,
\end{eqnarray}
\begin{eqnarray}
\beta_{1,m,\sigma}=g_\mathrm{AM}\Psi_{\sigma} \Phi_{m}^\ast,
\end{eqnarray}
\begin{eqnarray}
\beta_{2,m,\sigma}=g_\mathrm{AM}\Psi_{\sigma} \Phi_{m},
\end{eqnarray}
\begin{eqnarray}
\beta_{3,m,\sigma}=g_\mathrm{AM}\Psi_{\sigma}^\ast \Phi_{m}^\ast,
\end{eqnarray}
\begin{eqnarray}
\beta_{4,m,\sigma}=g_\mathrm{AM}\Psi_{\sigma}^\ast \Phi_{m},
\end{eqnarray}
\begin{eqnarray}
\alpha_{m,\sigma,\textbf{k}}=\overline{g}\sqrt{4\pi}\Psi_{\sigma, \textbf{Q}_\sigma}(\textbf{Q}_\sigma-\frac{\textbf{k}}{2})^2 Y_2^m(\widehat{\textbf{k}-2\textbf{Q}_\sigma}),
\end{eqnarray}
where $\epsilon_k=\frac{k^2}{2m}, \overline{1}=2$ and $\overline{2}=1$. We will diagonalize the Hamiltonian theoretically up to the order of $k^2$, and use exact diagonalization to testify our analysis in the meanwhile.
\\
\\

\subsection{Atomic Superfluid}
In the ASF phase, it has been found that the molecular modes are gapped, the relative mean-field $\Phi_m=0$, and the atoms are condensed at zero momentum $\textbf{Q}=0$. The full Bogliubov Hamiltonian is rewritten by substituting these conditions in
\begin{eqnarray}
H_f=\sum_{\textbf{k},i,j} \hat{c}_{i,\textbf{k}}^\dagger h_{\textbf{k}}^{i,j} \hat{c}_{j,\textbf{k}},
\end{eqnarray}
\\*
where $\hat{c}_{\textbf{k}}$ and $h_{\textbf{k}}$ are defined below,
\begin{widetext}
	\setcounter{MaxMatrixCols}{20}
	\newcommand{\colvec}[2][.8]{%
		\scalebox{#1}{%
			\renewcommand{\arraystretch}{.8}%
			$\begin{bmatrix}#2\end{bmatrix}$%
		}
	}
	\begin{eqnarray}
	\hat{c}_{\textbf{k}}^\dagger=\colvec[.8]{\hat{a}_{1,\textbf{k}}^\dagger  & \hat{a}_{1,-\textbf{k}} & \hat{a}_{2,\textbf{k}}^\dagger & \hat{a}_{2,-\textbf{k}}  & \hat{b}_{-2,\textbf{k}}^\dagger &  \hat{b}_{2,-\textbf{k}} & \hat{b}_{-1,\textbf{k}}^\dagger & \hat{b}_{1,-\textbf{k}} & \hat{b}_{0,\textbf{k}}^\dagger & \hat{b}_{0,-\textbf{k}} &  \hat{b}_{1,\textbf{k}}^\dagger & \hat{b}_{-1,-\textbf{k}} &  \hat{b}_{2,\textbf{k}}^\dagger & \hat{b}_{-2,-\textbf{k}}},
	\end{eqnarray}

	\begin{eqnarray}
	h_{\textbf{k}}=\colvec[.7] {\varepsilon_{1,\textbf{k}} & 2\tilde{\lambda}_{1}^\ast & t_1   & t_{2,\textbf{k}}^\ast & -\alpha_{-2,2,\textbf{k}}^\ast & 0 & -\alpha_{-1,2,\textbf{k}}^\ast & 0 & -\alpha_{0,2,\textbf{k}}^\ast &  0 & -\alpha_{1,2,\textbf{k}}^\ast & 0&-\alpha_{2,2,\textbf{k}}^\ast&0 \\
		2\tilde{\lambda}_{1} & \varepsilon_{1,-\textbf{k}} & t_{2,-\textbf{k}}  & t_1^\ast &  0&-\alpha_{2,2,-\textbf{k}}&0&-\alpha_{1,2,-\textbf{k}}&0& -\alpha_{0,2,-\textbf{k}}& 0 & -\alpha_{-1,2,-\textbf{k}} & 0 & -\alpha_{-2,2,-\textbf{k}} \\
		t_1^\ast& t_{2,-\textbf{k}}^\ast & \varepsilon_{2,\textbf{k}}  & 2\tilde{\lambda}_{2}^\ast   & -\alpha_{-2,1,\textbf{k}}^\ast & 0 & -\alpha_{-1,1,\textbf{k}}^\ast & 0 & -\alpha_{0,1,\textbf{k}}^\ast&  0 & -\alpha_{1,1,\textbf{k}}^\ast & 0&-\alpha_{2,1,\textbf{k}}^\ast&0\\
		t_{2,\textbf{k}}& t_{1}& 2\tilde{\lambda}_2 & \varepsilon_{2,-\textbf{k}} & 0&-\alpha_{2,1,-\textbf{k}}&0&-\alpha_{1,1,-\textbf{k}}&0& -\alpha_{0,1,-\textbf{k}}& 0 & -\alpha_{-1,1,-\textbf{k}} & 0 & -\alpha_{-2,1,-\textbf{k}} \\
		-\alpha_{-2,2,\textbf{k}}& 0 & -\alpha_{-2,1,\textbf{k}}  & 0 & \omega_{-2,k} & 0 & 0 & 0 & 0 & 0 & 0 & 0 & 0 & 0\\
		0& -\alpha_{2,2,-\textbf{k}}^\ast & 0  & -\alpha_{2,1,-\textbf{k}}^\ast & 0 & \omega_{2,k} & 0 & 0 & 0 & 0 & 0 & 0 & 0 & 0\\
		-\alpha_{-1,2,\textbf{k}}& 0 & -\alpha_{-1,1,\textbf{k}}  & 0 & 0 & 0 & \omega_{-1,k} & 0 & 0 & 0 & 0 & 0 & 0 & 0\\
		0& -\alpha_{1,2,-\textbf{k}}^\ast & 0  & -\alpha_{1,1,-\textbf{k}}^\ast & 0 & 0 & 0 & \omega_{1,k} & 0 & 0 & 0 & 0 & 0\\
		-\alpha_{0,2,\textbf{k}}& 0 & -\alpha_{0,1,\textbf{k}}  & 0 & 0 & 0 & 0& 0& \omega_{0,k} & 0 & 0 & 0 & 0 & 0\\
		0& -\alpha_{0,2,-\textbf{k}}^\ast & 0 & -\alpha_{0,1,-\textbf{k}}^\ast & 0 & 0 & 0 & 0& 0 & \omega_{0,k}  & 0 & 0 & 0 & 0\\
		-\alpha_{1,2,\textbf{k}} & 0 & -\alpha_{1,1,\textbf{k}} & 0 & 0 & 0 & 0& 0& 0 & 0& \omega_{1,k} & 0 & 0 & 0\\
		0 & -\alpha_{-1,2,-\textbf{k}}^\ast & 0 & -\alpha_{-1,1,-\textbf{k}}^\ast & 0 & 0 & 0 & 0& 0& 0 & 0 & \omega_{-1,k}  & 0 & 0\\
		-\alpha_{2,2,\textbf{k}} & 0 & -\alpha_{2,1,\textbf{k}} & 0 & 0 & 0 & 0& 0& 0 & 0& 0 & 0& \omega_{2,k} & 0\\
		0 & -\alpha_{-2,2,-\textbf{k}}^\ast & 0 & -\alpha_{-2,1,-\textbf{k}}^\ast & 0 & 0 & 0 & 0& 0& 0 & 0& 0 & 0 & \omega_{-2,k}},
	\end{eqnarray}
\end{widetext}
\noindent
where the reduced parameters are defined below,
\begin{eqnarray}
\varepsilon_{\sigma,\textbf{k}}=\epsilon_k-\mu_\sigma+2\lambda_{\sigma,\sigma}|\Psi_{\sigma}|^2+\frac{\lambda_{12}+\lambda_{21}}{2}|\Psi_{\overline{\sigma}}|^2,
\end{eqnarray}
\begin{eqnarray}
\omega_{k}=\frac{1}{2}\epsilon_{\textbf{k}}+z(\frac{1}{2}\epsilon_{\textbf{k}})^2-\mu_M+g_\mathrm{AM}(|\Psi_1|^2+|\Psi_2|^2),
\end{eqnarray}
\begin{eqnarray}
\tilde{\lambda}_\sigma=\frac{1}{2}\lambda_{\sigma,\sigma}\Psi_\sigma^{\ast2},
\end{eqnarray}
\begin{eqnarray}
t_1=\frac{1}{2}(\lambda_{12}+\lambda_{21})\Psi_1 \Psi_2^\ast,
\end{eqnarray}
\begin{eqnarray}
t_{2,\textbf{k}}=\frac{1}{2}(\lambda_{12}+\lambda_{21})\Psi_1^\ast \Psi_2^\ast,
\end{eqnarray}
\begin{eqnarray}
\alpha_{m,\sigma,\textbf{k}}=\frac{1}{4}\overline{g}\sqrt{4\pi}\Psi_{\sigma, \textbf{Q}_\sigma}k^2 Y_2^m(\hat{\textbf{k}}).
\end{eqnarray}

\begin{figure}[tbp]
	\includegraphics[width=7.5cm]{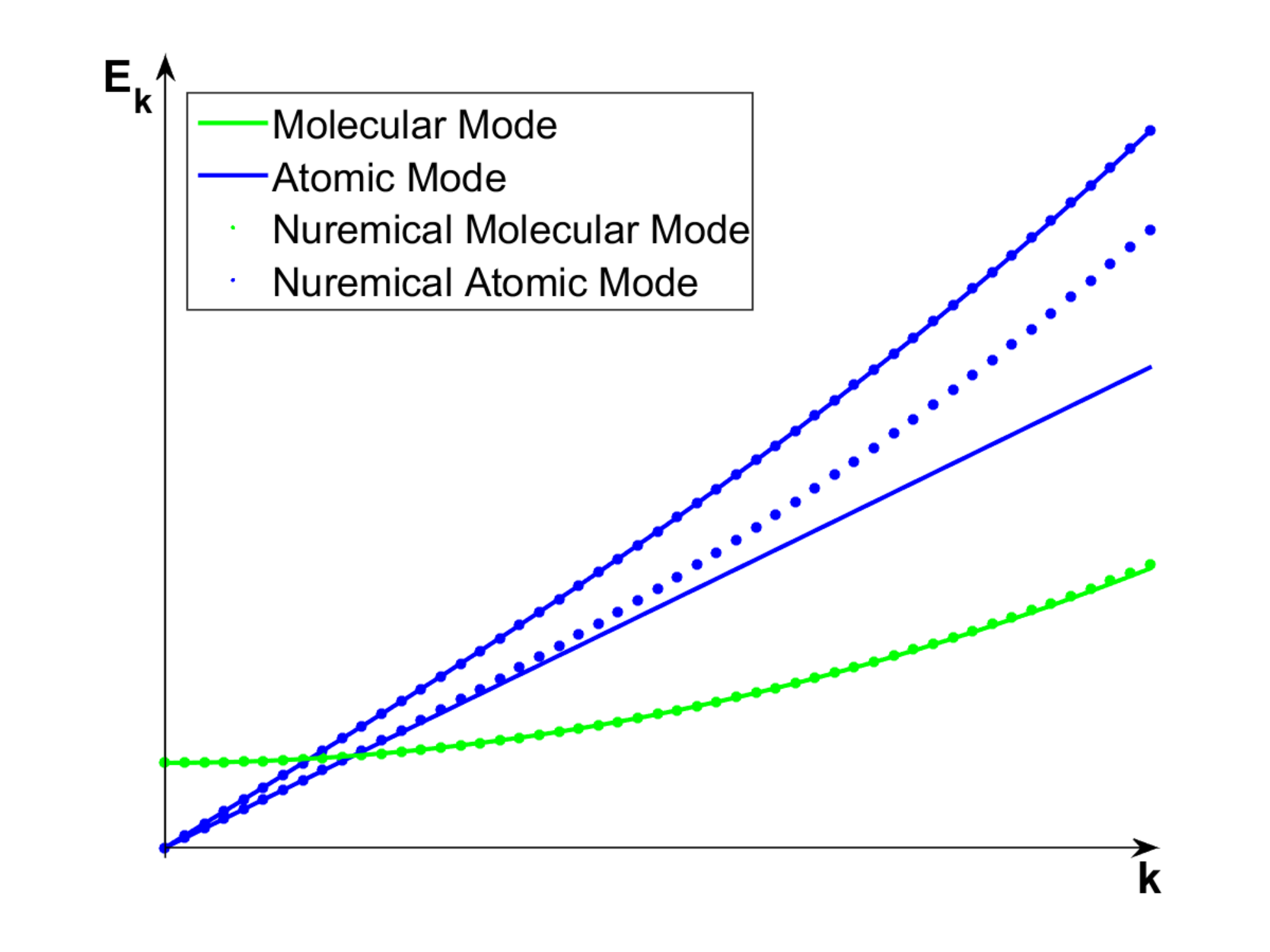}
	\caption{Schematic ASF phase excitation spectrum. All the molecular modes are gapped, but the atomic modes are gapless. The five molecule modes are degenerate. The numerical results and theoretical results fit well in small $k$ regime.} 
	\label{fig:model}
\end{figure}

To find out the atomic modes, we need to integrate the molecular modes out (see supplementary). In the low energy regime, when $k\to0$, we consider the dispersion up to $k^2$ order. The atomic modes are given by
\begin{eqnarray}
E_{1,k}^A=\sqrt{\frac{k^2}{2m}(\frac{k^2}{2m}+2 \lambda n_A)},
\end{eqnarray}
\begin{eqnarray}
E_{2,k}^A=\sqrt{\frac{(2\lambda-\lambda_{12}-\lambda_{21})n_A}{2m}}k,
\end{eqnarray}
\noindent
where $n_A$ is atom condensate density, $n_A=|\Psi_1|^2+|\Psi_2|^2$. Obviously, the atomic modes are gapless excitaitons present in the superfluid states. The molecular modes are given by
\begin{eqnarray}
E_{-2,-1,0,1,2,k}^M=\frac{k^2}{4m}+\nu-2\lambda n_A +g_\mathrm{AM}n_A.
\end{eqnarray}
\\*
The molecular modes dispersion has energy gap $\nu-2\lambda n_A +g_\mathrm{AM}n_A$. When it vanishes, we have the transition from ASF phase to AMSF phase at the detuning value
\begin{eqnarray}
\nu=2\lambda n_A - g_\mathrm{AM}n_A,
\end{eqnarray}
which is consistent with Eq.~(\ref{eq_nu2d})(we have $n_A=\mu/\lambda$). Fig. 4 shows the theoretical results and numerical results. They fit well in the small $k$ region.

\subsection{Molecular Superfluid}
In the MSF phase, the atomic modes are gapped and their mean-fields $\Psi_1=\Psi_2=0$. The vanishing of atomic mean-fields results in $\alpha_{m,\sigma,\textbf{k}}=0$, which means the atomic Hamiltonian and molecular Hamiltonian are separable. Thus we can obtain the atomic modes and molecular modes separately without considering their interactions.
\begin{eqnarray}
H_f=H_A+H_M,
\end{eqnarray}
\\*
Atomic Hamiltonian $H_A$ is given as,
\begin{eqnarray}
\newcommand{\colvec}[2][.8]{%
	\scalebox{#1}{%
		\renewcommand{\arraystretch}{.8}%
		$\begin{bmatrix}#2\end{bmatrix}$%
	}
}
H_A=\sum_{k}\colvec[.8] {\hat{a}_{1,\textbf{k}}^\dagger \\ \hat{a}_{2,\textbf{k}}^\dagger\\ \hat{a}_{1,-\textbf{k}}  \\ \hat{a}_{2,-\textbf{k}}}^T
\colvec[.8] {\varepsilon_{1,\textbf{k}} & 0 & 0 & t_{2,\textbf{k}}^\ast\\
	0 & \varepsilon_{2,\textbf{k}} & t_{2,-\textbf{k}}^\ast & 0\\
	0 & t_{2,-\textbf{k}} & \varepsilon_{1,-\textbf{k}} & 0\\
	t_{2,\textbf{k}}&0&0 & \varepsilon_{2,-\textbf{k}}}
\colvec[.8] {\hat{a}_{1,\textbf{k}} \\ \hat{a}_{2,\textbf{k}} \\ \hat{a}_{1,-\textbf{k}}^\dagger \\  \hat{a}_{2,-\textbf{k}}^\dagger},
\end{eqnarray}
\\*
where the reduced parameters are defined below,
\begin{eqnarray}
\varepsilon_{\sigma,\textbf{k}}=\epsilon_{\textbf{k}}-\mu_\sigma+g_\mathrm{AM} n_M,
\end{eqnarray}
\begin{eqnarray}
t_{2,\textbf{k}}=\overline{g}\sqrt{4\pi}k^2\sum_{m}\Phi_m^\ast Y_2^m(\hat{\textbf{k}}).
\end{eqnarray}
\noindent
The respective atomic dispersions are degenerate,
\newcommand*{\Scale}[2][4]{\scalebox{#1}{$#2$}}
\begin{eqnarray}\label{eq_MSFdisp}
\Scale[1]{
	E_{k}^A=\sqrt{\varepsilon_{\textbf{k}}^2-4\pi\overline{g}^2\textbf{k}^4\sum_{m,n}\Phi_m Y_2^m(\widehat{\textbf{k}})\Phi_n^\ast Y_2^n(\widehat{\textbf{k}})},}
\end{eqnarray}
\\*
For the MSF phase, $\mu = \frac{1}{2}(\mu_M+\nu)<\frac{1}{2}(\mu_M+\nu_1^d)$, from which we can obtain $-\mu+g_\mathrm{AM}n_M>0$. So in Eq. (\ref{eq_MSFdisp}) $k=0$ gives us the energy gap,
\begin{eqnarray}
\Delta E_{k}^A=-\mu+g_\mathrm{AM}n_M,
\end{eqnarray}
\noindent
By setting $\Delta E_{k}^a=0$, the atomic modes become gapless and the atomic condensates are emergent, which gives us
\begin{eqnarray}
\mu=g_\mathrm{AM}n_M.
\end{eqnarray}
\noindent
Applying $\nu=2\mu-\mu_M$ and $\mu_M=g_0 n_M$(see Eq.~(\ref{eq_GSmol})), we obtain the transition from the MSF to AMSF phase at the detuning value,
\begin{eqnarray}
\nu=(2-\frac{g_0}{g_\mathrm{AM}})\mu,
\end{eqnarray}
\\*
which is consistent with Eq.~(\ref{eq_nu1d}).

\begin{figure}[tbp]
	\includegraphics[width=7.5cm]{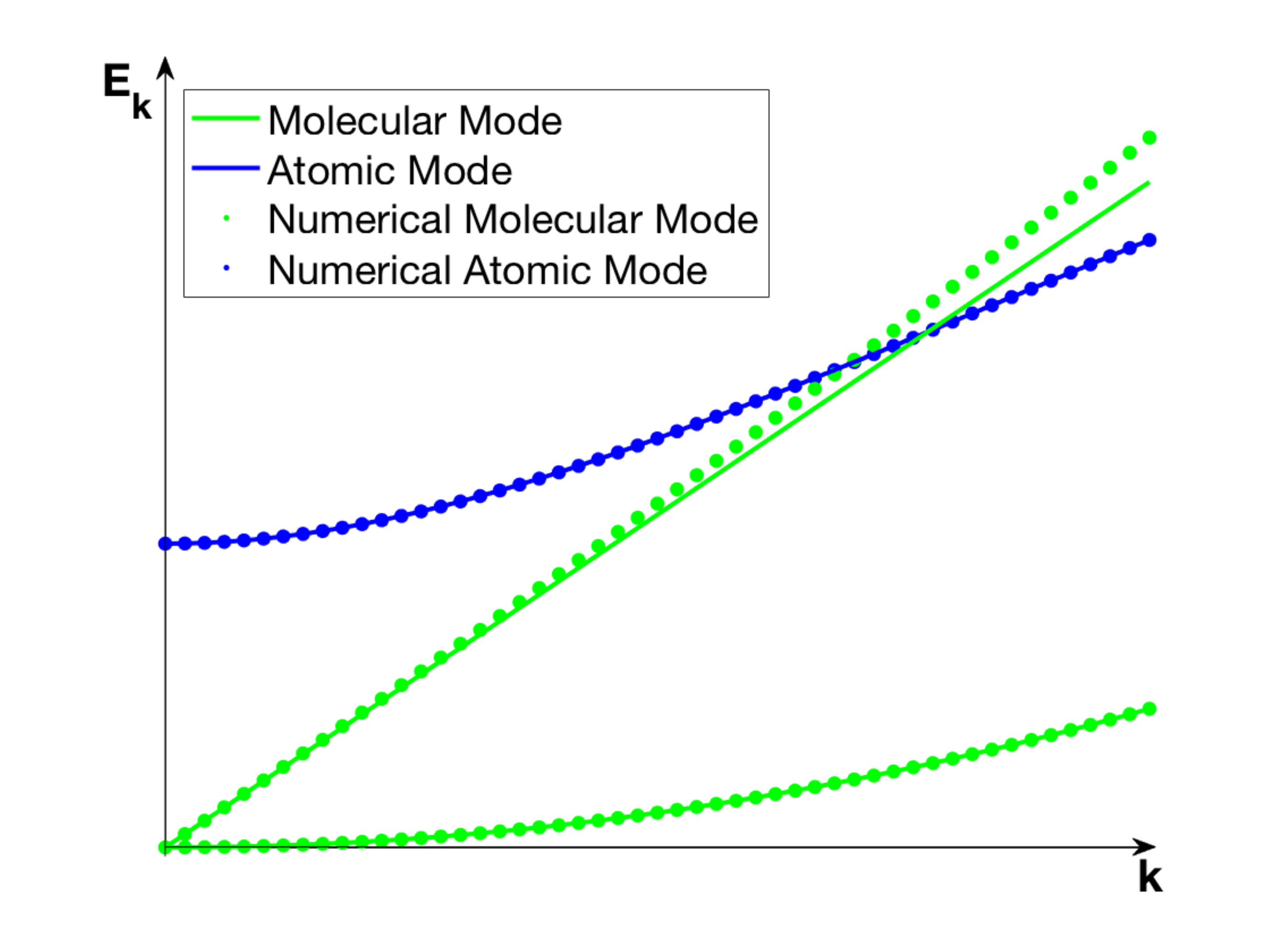}
	\caption{Schematic MSF phase excitation spectrum. The atomic modes are gapped and degenerate. All the molecular modes are gapless, $m=\pm 1,\pm 2$ are degenerate on the lower green line, $m=0$ is on the upper green line.} 
	\label{fig:model}
\end{figure}

Let us turn to the molecular modes in the MSF phase. Referring to the mean-field ground state~(\ref{eq_GSmol}), we choose the simplest case $D=1$ to explore its low energy excitation,
\begin{eqnarray}
\Phi_{-2,-1,1,2}=0,\\
\Phi_{0}=\sqrt{n_M}.
\end{eqnarray}
\noindent
The molecular Hamiltonian is hence given as
\begin{eqnarray}
H_m=\sum_{\textbf{k},i,j} \hat{b}_{i,\textbf{k}}^\dagger h_{m,\textbf{k}}^{i,j} \hat{b}_{j,\textbf{k}},
\end{eqnarray}
\\*
where $\hat{b}_{i,\textbf{k}}$ and $h_{m,\textbf{k}}$ are defined below,
\begin{widetext}
	\begin{eqnarray}
	\hat{b}_{i,\textbf{k}}^\dagger=(\hat{b}_{-2,\textbf{k}}^\dagger,\hat{b}_{2,-\textbf{k}},\hat{b}_{-1,\textbf{k}}^\dagger,\hat{b}_{1,-\textbf{k}},\hat{b}_{0,\textbf{k}}^\dagger,\hat{b}_{0,-\textbf{k}},\hat{b}_{1,\textbf{k}}^\dagger,\hat{b}_{-1,-\textbf{k}},\hat{b}_{2,\textbf{k}}^\dagger,\hat{b}_{-2,-\textbf{k}}),
	\end{eqnarray}
	\setcounter{MaxMatrixCols}{20}
	\newcommand{\colvec}[2][.8]{%
		\scalebox{#1}{%
			\renewcommand{\arraystretch}{.8}%
			$\begin{bmatrix}#2\end{bmatrix}$%
		}
	}
	\begin{eqnarray}\label{eq_hamibogo}
	h_{\textbf{k}}=\begin{pmatrix} \Omega_{-2} &0&0&0&0\\
	0&\Omega_{-1} &0&0&0\\0&0&\Omega_{0} &0&0\\
	0&0&0&\Omega_{1} &0\\0&0&0&0&\Omega_{2}\end{pmatrix},
	\end{eqnarray}
\end{widetext}
\noindent
where $\Omega_{n}$ is a $2\times 2$ matrix, defined as
\begin{eqnarray}
\Omega_{n \ne 0}=\begin{pmatrix}
\omega_{n,k} & 0 \\ 0 & \omega_{-n,k}
\end{pmatrix},
\end{eqnarray}
\begin{eqnarray}
\Omega_{n = 0}=\begin{pmatrix}
\omega_{0,k} & 2\delta_0^\ast \\ 2\delta_0 & \omega_{0,k}
\end{pmatrix},
\end{eqnarray}
\\*
the reduced parameters are
\begin{eqnarray}
\omega_{n \ne 0,k}&=&\frac{1}{2}\epsilon_{\textbf{k}}+z(\frac{1}{2}\epsilon_{\textbf{k}})^2-\mu_M
+
g_{0}n_M,
\end{eqnarray}
\begin{eqnarray}
\omega_{n=0,k}&=&\frac{1}{2}\epsilon_{\textbf{k}}+z(\frac{1}{2}\epsilon_{\textbf{k}})^2-\mu_M+
2g_0 n_M,
\end{eqnarray}
\begin{eqnarray}
\delta_0=\frac{1}{2}g_0 n_M.
\end{eqnarray}

The diagonalization of Hamiltonian~(\ref{eq_hamibogo}) is straightforward. Up to $k^2$ order, the molecular mode dispersions are
\begin{eqnarray}
E_{0,k}^M=\sqrt{\frac{g_0n_M}{2m}}k,
\end{eqnarray}
\begin{eqnarray}
E_{n=\pm 1,\pm 2,k}^M=\frac{k^2}{4m},
\end{eqnarray}
\\*
the five molecular modes are all gapless, which proves that they are in superfluid state. Fig. 5 shows the consistency between the theoretical results and numerical results.
\\
\\

\subsection{Atomic-Molecular Superfluid}
For the intermediate phase, both atomic and molecular condensates exist. Hence, they define a complicated coupled Hamiltonian, Eq.~(\ref{BogoH}). The molecular and atomic condensate mean-field solutions are given as
\begin{eqnarray}
\newcommand{\colvec}[2][.8]{%
	\scalebox{#1}{%
		\renewcommand{\arraystretch}{.8}%
		$\begin{bmatrix}#2\end{bmatrix}$%
	}
}
\begin{split}
\Phi&=\sqrt{\frac{g_\mathrm{AM}\mu-\lambda\mu_M}{g_\mathrm{AM}^2-g_0\lambda}}D (0,0,1,0,0)^T,
\end{split}
\end{eqnarray}

\begin{eqnarray}
\Psi_{1,2}=\sqrt{\frac{g_0\mu-g_\mathrm{AM}\mu_M}{2\lambda g_0-2g_\mathrm{AM}^2}}.
\end{eqnarray}
\noindent
Similar to what we have achieved in the MSF phase, we choose the simplest case to compute the spectrums, $D=1$, so that the Hamiltonian is reduced to
\begin{eqnarray}
H_f=\sum_{\textbf{k},i,j} \hat{c}_{i,\textbf{k}}^\dagger h_{\textbf{k}}^{i,j} \hat{c}_{j,\textbf{k}},
\end{eqnarray}
\\*
where $\hat{c}_{\textbf{k}}$ and $h_{\textbf{k}}$ are defined below,
\begin{widetext}
	\setcounter{MaxMatrixCols}{20}
	\newcommand{\colvec}[2][.8]{%
		\scalebox{#1}{%
			\renewcommand{\arraystretch}{.8}%
			$\begin{bmatrix}#2\end{bmatrix}$%
		}
	}
	\begin{eqnarray}
	\hat{c}_{\textbf{k}}^\dagger=\colvec[.8]{\hat{a}_{1,\textbf{k}}^\dagger  & \hat{a}_{1,-\textbf{k}} & \hat{a}_{2,\textbf{k}}^\dagger & \hat{a}_{2,-\textbf{k}}  & \hat{b}_{-2,\textbf{k}}^\dagger &  \hat{b}_{2,-\textbf{k}} & \hat{b}_{-1,\textbf{k}}^\dagger & \hat{b}_{1,-\textbf{k}} & \hat{b}_{0,\textbf{k}}^\dagger & \hat{b}_{0,-\textbf{k}} &  \hat{b}_{1,\textbf{k}}^\dagger & \hat{b}_{-1,-\textbf{k}} &  \hat{b}_{2,\textbf{k}}^\dagger & \hat{b}_{-2,-\textbf{k}}},
	\end{eqnarray}
	
	\begin{eqnarray}
	h_{\textbf{k}}^{i,j}=\colvec[.53] {\varepsilon_{1,\textbf{k}} & 2\tilde{\lambda}_{1}^\ast & t_1   & t_{2,\textbf{k}}^\ast & -\alpha_{-2,2,\textbf{k}}^\ast & 0 & -\alpha_{-1,2,\textbf{k}}^\ast & 0 & -\alpha_{0,2,\textbf{k}}^\ast+\beta_{1,0,1} &  \beta_{2,0,1} & -\alpha_{1,2,\textbf{k}}^\ast & 0&-\alpha_{2,2,\textbf{k}}^\ast&0 \\
		2\tilde{\lambda}_{1} & \varepsilon_{1,-\textbf{k}} & t_{2,-\textbf{k}}  & t_1^\ast &  0&-\alpha_{2,2,-\textbf{k}}&0&-\alpha_{1,2,-\textbf{k}}&\beta_{3,0,1}& -\alpha_{0,2,-\textbf{k}}+\beta_{4,0,1} & 0 & -\alpha_{-1,2,-\textbf{k}} & 0 & -\alpha_{-2,2,-\textbf{k}} \\
		t_1^\ast& t_{2,-\textbf{k}}^\ast & \varepsilon_{2,\textbf{k}}  & 2\tilde{\lambda}_{2}^\ast   & -\alpha_{-2,1,\textbf{k}}^\ast & 0 & -\alpha_{-1,1,\textbf{k}}^\ast & 0 & -\alpha_{0,1,\textbf{k}}^\ast+\beta_{1,0,2} &  \beta_{2,0,2} & -\alpha_{1,1,\textbf{k}}^\ast & 0&-\alpha_{2,1,\textbf{k}}^\ast&0\\
		t_{2,\textbf{k}}& t_{1}& 2\tilde{\lambda}_2 & \varepsilon_{2,-\textbf{k}} & 0&-\alpha_{2,1,-\textbf{k}}&0&-\alpha_{1,1,-\textbf{k}}&\beta_{3,0,2}& -\alpha_{0,1,-\textbf{k}}+\beta_{4,0,2} & 0 & -\alpha_{-1,1,-\textbf{k}} & 0 & -\alpha_{-2,1,-\textbf{k}} \\
		-\alpha_{-2,2,\textbf{k}}& 0 & -\alpha_{-2,1,\textbf{k}}  & 0 & \omega_{-2,k} & 0 & 0 & 0 & 0 & 0 & 0 & 0 & 0 & 0\\
		0& -\alpha_{2,2,-\textbf{k}}^\ast & 0  & -\alpha_{2,1,-\textbf{k}}^\ast & 0 & \omega_{2,k} & 0 & 0 & 0 & 0 & 0 & 0 & 0 & 0\\
		-\alpha_{-1,2,\textbf{k}}& 0 & -\alpha_{-1,1,\textbf{k}}  & 0 & 0 & 0 & \omega_{-1,k} & 0 & 0 & 0 & 0 & 0 & 0 & 0\\
		0& -\alpha_{1,2,-\textbf{k}}^\ast & 0  & -\alpha_{1,1,-\textbf{k}}^\ast & 0 & 0 & 0 & \omega_{1,k} & 0 & 0 & 0 & 0 & 0\\
		-\alpha_{0,2,\textbf{k}}+\beta_{1,0,1}^\ast& \beta_{3,0,1}^\ast & -\alpha_{0,1,\textbf{k}}+\beta_{1,0,2}^\ast  & \beta_{3,0,2}^\ast & 0 & 0 & 0& 0& \omega_{0,k} & 2\delta_{0}^\ast & 0 & 0 & 0 & 0\\
		\beta_{2,0,1}^\ast& -\alpha_{0,2,-\textbf{k}}^\ast+\beta_{4,0,1}^\ast & \beta_{2,0,2}^\ast & -\alpha_{0,1,-\textbf{k}}^\ast+\beta_{4,0,2}^\ast & 0 & 0 & 0 & 0& 2\delta_{0} & \omega_{0,k}  & 0 & 0 & 0 & 0\\
		-\alpha_{1,2,\textbf{k}} & 0 & -\alpha_{1,1,\textbf{k}} & 0 & 0 & 0 & 0& 0& 0 & 0& \omega_{1,k} & 0 & 0 & 0\\
		0 & -\alpha_{-1,2,-\textbf{k}}^\ast & 0 & -\alpha_{-1,1,-\textbf{k}}^\ast & 0 & 0 & 0 & 0& 0& 0 & 0 & \omega_{-1,k}  & 0 & 0\\
		-\alpha_{2,2,\textbf{k}} & 0 & -\alpha_{2,1,\textbf{k}} & 0 & 0 & 0 & 0& 0& 0 & 0& 0 & 0& \omega_{2,k} & 0\\
		0 & -\alpha_{-2,2,-\textbf{k}}^\ast & 0 & -\alpha_{-2,1,-\textbf{k}}^\ast & 0 & 0 & 0 & 0& 0& 0 & 0& 0 & 0 & \omega_{-2,k}},
	\end{eqnarray}
\end{widetext}
where the reduced parameters are given as
\begin{eqnarray}
\varepsilon_{\sigma,\textbf{k}}&=&\left.\nonumber\epsilon_{\textbf{k}}-\mu+\lambda_{\sigma,\sigma}n_A+\frac{1}{4}(\lambda_{12}+\lambda_{21})n_A\right.\\
&& \left.
+
g_\mathrm{AM}n_M\right.,
\end{eqnarray}
\begin{eqnarray}
\omega_{m,k}&=&\left.\frac{1}{2}\epsilon_{\textbf{k}}+z(\frac{1}{2}\epsilon_{\textbf{k}})^2-\mu_M+g_{0}n_M\right.\\
&& \left.\nonumber
+
g_{0}n_M \delta_{m,0}+g_\mathrm{AM}n_A\right.,
\end{eqnarray}
\begin{eqnarray}
\tilde{\lambda}_\sigma=\frac{1}{4}\lambda_{\sigma,\sigma}n_A,
\end{eqnarray}
\begin{eqnarray}
t_1=\frac{1}{4}(\lambda_{12}+\lambda_{21})n_A,
\end{eqnarray}
\begin{eqnarray}
t_{2,\textbf{k}}=\frac{1}{4}(\lambda_{12}+\lambda_{21})n_A-\overline{g}\sqrt{4\pi n_M}k^2 Y_2^0(\hat{\textbf{k}}),
\end{eqnarray}
\begin{eqnarray}
\delta_m=\frac{1}{2}g_0 n_M \delta_{m,0},
\end{eqnarray}
\begin{eqnarray}
\beta_{i,0,\sigma}=g_\mathrm{AM}\sqrt{\frac{n_A n_M}{2}}, i=1,2,3,4,
\end{eqnarray}
\begin{eqnarray}
\alpha_{m,\sigma,\textbf{k}}=\frac{\sqrt{2\pi}}{4}\overline{g}\sqrt{n_A}k^2 Y_2^m(\hat{\textbf{k}}),
\end{eqnarray}
\noindent
where $\delta_{m,0}$ is Kronecker delta function, and $n_A, n_M$ represent the atom and molecule condensate density respectively.

\begin{figure}[tbp]
	\includegraphics[width=7.5cm]{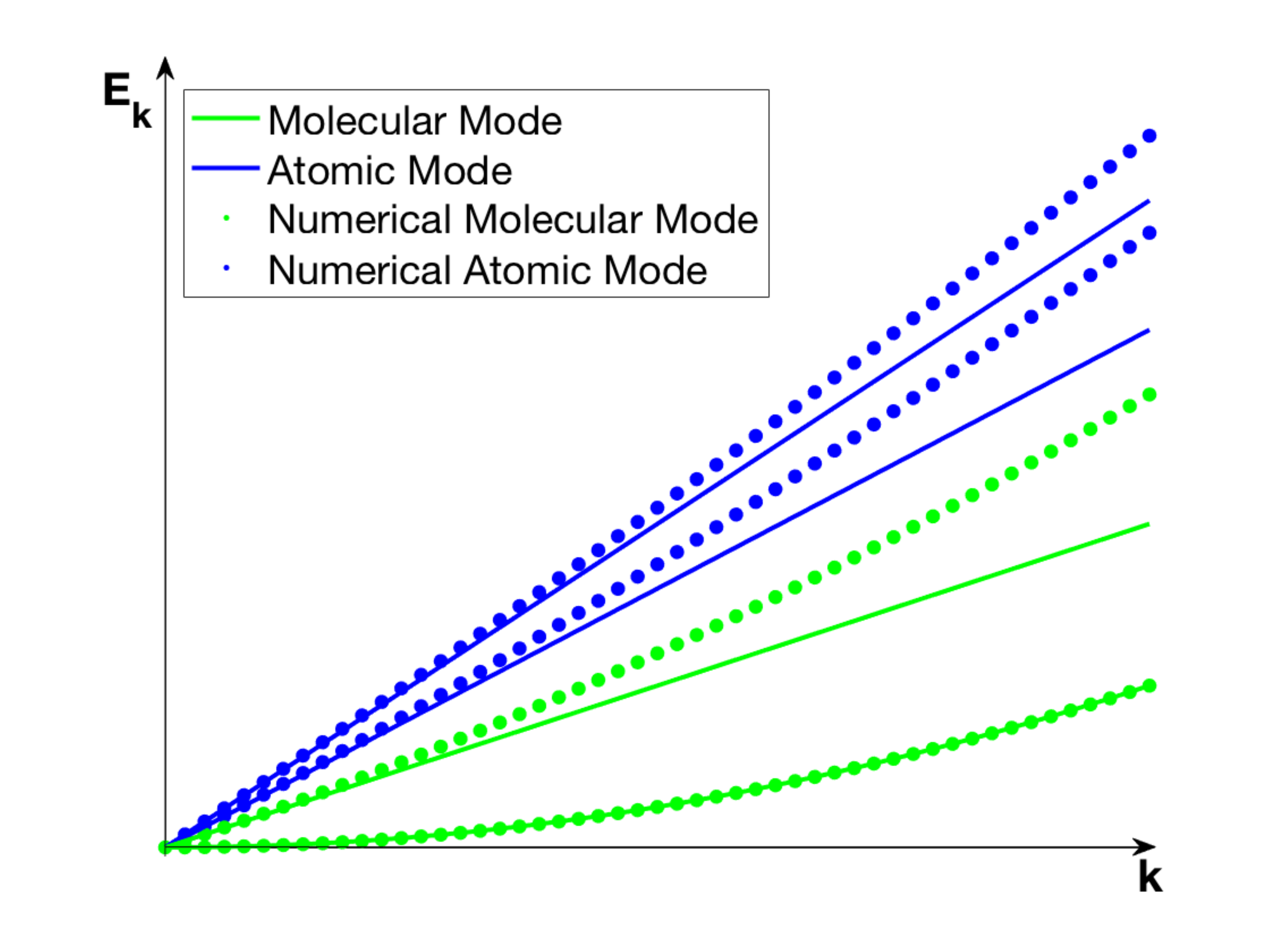}
	\caption{Schematic AMSF phase excitation spectrum. The atomic modes are gapless on the two blue lines. The molecular modes are also gapless: $m=\pm 1,\pm 2$ are degenerate on the lower green line; $m=0$ is on the upper green line.} 
	\label{fig:model}
\end{figure}

Diagonalizing this Hamiltonian leads to the spectrums up to the order of $k$,
\begin{eqnarray}
E_{1,k}^A=\sqrt{\frac{\lambda n_A}{m}}k,
\end{eqnarray}
\begin{eqnarray}
E_{2,k}^A=\sqrt{(2\lambda-\lambda_{12}-\lambda_{21})n_A (\frac{1}{2m}-\sqrt{5n_M}\overline{g})}k,
\end{eqnarray}
\begin{eqnarray}
E_{0,k}^M=\sqrt{\frac{(g_0+\frac{g_\mathrm{AM}^2}{\lambda})n_M}{2m}}k,
\end{eqnarray}
\begin{eqnarray}
E_{n=\pm 1,\pm 2,k}^M=\frac{k^2}{4m},
\end{eqnarray}

Fig. 6 shows the consistency between the theoretical results and numerical results.

\section{Conclusion}
In this paper, we study the mean-field ground state of a $d$-wave interacting Bose gas, and it is found that there are three superfluid phases: atomic, molecular and atomic-molecular superfluid phases. What is most surprising is that unlike the $p$-wave case~\cite{radzihovsky2009p,pwave2011Sungsoo,spontaneous2019li}, we find the atomic superfluid does not carry finite momentum. Further, we study the low-energy excitation spectrum above the superfluid phases. Our work provides a basic reference for the experiment on degenerate $d$-wave interacting Bose gas.

\section*{Acknowledgments}
The authors are indebted to Bing Zhu and Chao Gao for helpful discussion. This work is supported by the AFOSR Grant No. FA9550-16-1-0006 and the MURI-ARO Grant No. W911NF17-1-0323 (Z. L. and W.V. L.), the National Natural Science Foundation of China  (Grant No. 11904228 and No. 11804221) and the National Postdoctoral Program for Innovative Talents of China (Grant No. BX201700156) (J.-S. P.), the Science and Technology Commission of Shanghai Municipality (Grants No.16DZ2260200) and National Natural Science Foundation of China (Grants No.11655002) (J.-S. P. and W.V. L.), and the Overseas Scholar Collaborative Program of NSF of China No. 11429402 sponsored by Peking University (W.V. L.).

\section{Supplementary}

\vspace{1cm}
\subsection{Atomic Order Parameters}
Up to the order of Q, the collinear states fall into two universality classes, FF-like~\cite{FF} and LO-like~\cite{LO} single harmonic forms,
\begin{eqnarray}
\nonumber
\Psi_{\sigma}^{FF}&=&
\Psi_{\sigma,\textbf{Q}_{\sigma}}e^{i \textbf{Q}_{\sigma} \cdot \textbf{r}},
\end{eqnarray}
\begin{eqnarray}
\nonumber
\Psi_{\sigma}^{LO}&=&
\Psi_{\sigma,\textbf{Q}}e^{i \textbf{Q} \cdot \textbf{r}}
+
\Psi_{\sigma,-\textbf{Q}}e^{-i \textbf{Q} \cdot \textbf{r}},
\end{eqnarray}
\noindent
where $\Psi_{\sigma,\textbf{Q}} = \Psi_{\sigma,-\textbf{Q}}$. We will prove in the following context that FF-like form is energetically preferred in a low energy regime.

In LO form, the free energy density is
\begin{eqnarray}
\nonumber
f&=&F/V=f_{M} + f_{\textbf{Q}},
\end{eqnarray}
where
\begin{eqnarray}
f_M&=&\left.\nonumber
\sum_{m=-2}^{2}-\mu_M|\Phi_m|^2
+
\sum_{m,n=-2}^{2}\frac{g_0}{2} (\Phi_m^\ast \Phi_m)(\Phi_n^\ast \Phi_n), \right.
\end{eqnarray}
\begin{eqnarray}
f_{\textbf{Q}}&=&\nonumber
\frac{1}{V}\int_V d^3r\left.
\sum_{\sigma=1,2}4\varepsilon_{\textbf{Q}}(|\Psi_{\sigma,\textbf{Q}}|^2\cos^2{(\textbf{Q} \cdot \textbf{r})})\right.\\
&& \left.\nonumber
-
[\Delta^\ast(\Psi_{1,\textbf{Q}}\Psi_{2,-\textbf{Q}} + \Psi_{1,-\textbf{Q}}\Psi_{2,\textbf{Q}})+c.c.]
\right.\\
&& \left.\nonumber
+
\sum_{\sigma,\sigma'=1,2}\frac{\lambda_{\sigma,\sigma'}}{2}{|\Psi_\sigma|}^2{|\Psi_{\sigma'}|}^2,\right.
\end{eqnarray}
\noindent
and $\varepsilon_{\textbf{Q}} = \frac{Q^2}{2m}-\mu+g_\mathrm{AM}n_M, \Delta = \sum_{m=-2}^{2}\overline{g}\sqrt{4\pi} Q^2 \Phi_m Y_2^m(\hat{\textbf{Q}})$.
\noindent
For the quadratic part, we rewrite it in the matrix formula,
\begin{eqnarray}\nonumber
\newcommand{\colvec}[2][.8]{%
	\scalebox{#1}{%
		\renewcommand{\arraystretch}{.8}%
		$\begin{bmatrix}#2\end{bmatrix}$%
	}
}
f_{\textbf{Q}}^0=\frac{1}{V}\int_V d^3r\colvec[.5] {\Psi_{1,\textbf{Q}}^\ast \\ \Psi_{1,-\textbf{Q}} \\ \Psi_{2,\textbf{Q}}^\ast \\ \Psi_{2,-\textbf{Q}}}^T
\colvec[.5] {2\varepsilon_{\textbf{Q}}\cos^2{(\textbf{Q} \cdot \textbf{r})} & 0 & 0 & -\Delta_{\textbf{Q}}\\
	0 & 2\varepsilon_{\textbf{Q}}\cos^2{(\textbf{Q} \cdot \textbf{r})} & -\Delta_{\textbf{Q}}^\ast & 0\\
	0 & -\Delta_{\textbf{Q}} & 2\varepsilon_{\textbf{Q}}\cos^2{(\textbf{Q} \cdot \textbf{r})} & 0\\
	-\Delta_{\textbf{Q}}^\ast&0&0 & 2\varepsilon_{\textbf{Q}}\cos^2{(\textbf{Q} \cdot \textbf{r})}}
\colvec[.5] {\Psi_{1,\textbf{Q}} \\ \Psi_{1,-\textbf{Q}}^\ast \\ \Psi_{2,\textbf{Q}} \\ \Psi_{2,-\textbf{Q}}^\ast},
\end{eqnarray}
\noindent
We diagonalize the quadratic Hamiltonian, obtaining the eigenvector matrix and eigenvalues,
\begin{eqnarray}\nonumber
\newcommand{\colvec}[2][.8]{%
	\scalebox{#1}{%
		\renewcommand{\arraystretch}{.8}%
		$\begin{bmatrix}#2\end{bmatrix}$%
	}
}
U=\frac{1}{\sqrt{2}}
\colvec[.7] {e^{-i\theta_0} & 0 & -e^{-i\theta_0} & 0\\
	0 & e^{i\theta_0} & 0 & -e^{i\theta_0}\\
	0 & 1 & 0 & 1\\
	1&0&1 & 0},
\end{eqnarray}
\begin{eqnarray}\nonumber
\newcommand{\colvec}[2][.8]{%
	\scalebox{#1}{%
		\renewcommand{\arraystretch}{.8}%
		$\begin{bmatrix}#2\end{bmatrix}$%
	}
}
V=\colvec[.5] {2\varepsilon_{\textbf{Q}}\cos^2{(\textbf{Q} \cdot \textbf{r})}-|\Delta_{\textbf{Q}}|&0&0&0 \\ 0&2\varepsilon_{\textbf{Q}}\cos^2{(\textbf{Q} \cdot \textbf{r})}-|\Delta_{\textbf{Q}}|&0&0 \\ 0&0&2\varepsilon_{\textbf{Q}}\cos^2{(\textbf{Q} \cdot \textbf{r})}+|\Delta_{\textbf{Q}}|&0 \\ 0&0&0&2\varepsilon_{\textbf{Q}}\cos^2{(\textbf{Q} \cdot \textbf{r})}+|\Delta_{\textbf{Q}}|}.
\end{eqnarray}
\noindent
Hence, we can write the Nambu basis as
\begin{eqnarray}\nonumber
\newcommand{\colvec}[2][.8]{%
	\scalebox{#1}{%
		\renewcommand{\arraystretch}{.8}%
		$\begin{bmatrix}#2\end{bmatrix}$%
	}
}
\colvec[.7]{\Psi_{-,\textbf{Q}} \\ \Psi_{-,-\textbf{Q}}^\ast \\ \Psi_{+,\textbf{Q}} \\ \Psi_{+,-\textbf{Q}}^\ast}=\frac{1}{\sqrt{2}}\colvec[.7]{e^{i\theta_0}\Psi_{1,\textbf{Q}}+\Psi_{2,-\textbf{Q}}^\ast \\ e^{-i\theta_0}\Psi_{1,\textbf{Q}}^\ast+\Psi_{2,-\textbf{Q}} \\ -e^{i\theta_0}\Psi_{1,\textbf{Q}}+\Psi_{2,-\textbf{Q}}^\ast \\ -e^{-i\theta_0}\Psi_{1,\textbf{Q}}^\ast+\Psi_{2,-\textbf{Q}}}.
\end{eqnarray}

In the AMSF phase and ASF phase, the atoms prefer to stay at a lower energy level, such that in the ground state $\Psi_{+,\textbf{Q}}=0, \Psi_{+,-\textbf{Q}}^\ast=0$. We obtain $\Psi_{2,-\textbf{Q}}^\ast=e^{i\theta_0}\Psi_{1,\textbf{Q}}$, and $\Psi_{-,\textbf{Q}}=\sqrt{2}e^{i\theta_0}\Psi_{1,\textbf{Q}},\Psi_{-,-\textbf{Q}}=\sqrt{2}e^{-i\theta_0}\Psi_{1,\textbf{Q}}^\ast$, where $\theta_0$ is the angle of $\Delta$. The free energy can be rewritten in the form of the eigenvalues and eigenstates,
\begin{eqnarray}
E_{LO} &=&\left.\nonumber
 \frac{1}{V}\int_V d^3r 2(2\varepsilon_{\textbf{Q}}\cos^2{(\textbf{Q} \cdot \textbf{r})}-|\Delta_{\textbf{Q}}|)|\Psi_{-,\textbf{Q}}|^2\right.\\
 && \left.\nonumber
 +
 8\lambda |\Psi_{-,\textbf{Q}}|^2 \cos^4{(\textbf{Q} \cdot \textbf{r})}\right.,
\end{eqnarray}
\noindent
We use integral
\begin{eqnarray}\nonumber
\frac{1}{V}\int_V d^3r \cos^2{(\textbf{Q} \cdot \textbf{r})} = \frac{1}{2} + \frac{1}{2}\delta({\textbf{Q}}),
\end{eqnarray}
\begin{eqnarray}\nonumber
\frac{1}{V}\int_V d^3r \cos^4{(\textbf{Q} \cdot \textbf{r})} = \frac{3}{8} + \frac{5}{8}\delta({\textbf{Q}}),
\end{eqnarray}
\noindent
to obtain
\begin{eqnarray}
E_{LO} =\nonumber 2(\varepsilon_{\textbf{Q}}-|\Delta_{\textbf{Q}}|)|\Psi_{-,\textbf{Q}}|^2 + 3\lambda |\Psi_{-,\textbf{Q}}|^4,  \textbf{Q} \ne 0,
\end{eqnarray}
\begin{eqnarray}
E_{LO} =\nonumber 2(2\varepsilon_{\textbf{Q}}-|\Delta_{\textbf{Q}}|)|\Psi_{-,\textbf{Q}}|^2 + 8\lambda |\Psi_{-,\textbf{Q}}|^4,  \textbf{Q} = 0.
\end{eqnarray}
\noindent
Comparing with the calculations in the main context,
\begin{eqnarray}
E_{FF} =\nonumber (\varepsilon_{\textbf{Q}}-|\Delta_{\textbf{Q}}|)|\Psi_{-,\textbf{Q}}|^2 + \frac{1}{2}\lambda |\Psi_{-,\textbf{Q}}|^4.
\end{eqnarray}
\noindent
We can see that the FF-like state has lower energy, which is preferred in the ground state regime.

\subsection{Coupling Correction}
Suppose we have Bogliubov Hamiltonian for different modes in a block formula,
\begin{eqnarray}
\nonumber H=\begin{pmatrix}\xi^\dagger & \eta^\dagger\end{pmatrix}\begin{pmatrix}H_{11} & H_{12} \\ H_{21} & H_{22}\end{pmatrix}\begin{pmatrix}\xi \\ \eta\end{pmatrix}.
\end{eqnarray}
\noindent
To obtain the $\xi$ modes corrected by $\eta$ modes, we need to integrate out the $\eta$ modes,
\begin{eqnarray}
\nonumber \int d\eta d\eta^{\dagger}e^{-H}&=e^{-\xi^\dagger H_{11} \xi}\int d\eta d\eta^{\dagger} e^{-\eta^\dagger H_{22} \eta-(H_{21}\xi)^\dagger \eta-\eta^\dagger (H_{21}\xi)}\\
\nonumber &=\frac{1}{\det(H_{22})}e^{-\xi^\dagger H_{11} \xi+\xi^\dagger H_{12} H_{22}^{-1} H_{21} \xi},
\end{eqnarray}
\\*
thus, the corrected $\xi$ modes Hamiltonian is $H_{11} - H_{12} H_{22}^{-1} H_{21}$.

\bibliographystyle{apsrev4-1}
\bibliography{reference}




\end{document}